\documentstyle[epsf,aps]{revtex}

\begin{document}

\twocolumn[\hsize\textwidth\columnwidth\hsize\csname@twocolumnfalse\endcsname

\title{ Quasiparticle energy dispersion in doped two dimensional
quantum antiferromagnets}

\author{T. Xiang and J. M. Wheatley}

\address{ Research Center in Superconductivity, 
University of Cambridge, Madingley Road, 
Cambridge CB3 0HE, United Kingdom}

%\date{\today}

\maketitle

\begin{abstract}

The quasiparticle dispersion in the one-hole 
$t- t^\prime -t^{\prime \prime}-J $ model is studied.  Both the
finite-size diagonalization and the self-consistent Born
approximation calculations have been performed and compared.
The quasiparticle band structures in the hole and electron
doped high-$T_c$ cuprates are qualitatively different.  In
the hole doped compounds, the band maxima locate at $(\pm
\pi /2 , \pm \pi /2)$, while in the electron doped compounds
the band maxima locate at $(\pi, 0)$ and its equivalent
points.  The angle-resolved photoemission data for the 
quasiparticle dispersion of ${\rm Sr_2 Cu O_2 Cl_2}$
can be quantitatively reproduced using the one band
$t-t^\prime -t^{\prime \prime}- J $ model with the
three-site hopping term.

\end{abstract}


]

The energy dispersion of hole quasiparticles in the normal
states of two-dimensional CuO planes is of fundamental
interest for understanding the microscropic mechanism of
high temperature superconductivity.  Due to the strong
Coulomb repulsion between electrons, quasiparticles in
high-T$_c$ cuprates behave very differently from what
predicted by one-electron band calculations.  
Recently Wells {\it et al.}\cite{Wells95} has reported an
ARPES measurement on an insulating layered copper oxide
${\rm Sr_2CuO_2Cl_2}$. This is by
far the most direct measurement of the dispersion of a {\it
single hole} in an antiferromagnetic background since 
${\rm Sr_2CuO_2Cl_2}$ is difficult to dope. It
provides a direct test for the model Hamiltonians which are
proposed for high temperature superconductors.  Wells {\it
et al.}  has found that the the hole bandwidth and the
dispersion along the diagonal direction from $(0,0)$ to
$(\pi,\pi)$ in Brillion zone agree well with the
calculations based on the t-J model.  However, near $(\pi
,0)$ the data differ significantly from the prediction of
the t-J model and the dispersion from $(\pi , 0)$ to $(0,
\pi )$ is much greater than that in the t-J model.

Recently Nazarenko {\it et al.}\cite{Nazarenko95} has
calculated the hole dispersion in the one-band $t-t^\prime
-J$ model under the self-consistent Born approximation
(SCBA).  They found that the presence of the $t^\prime$-term
can enlarge the energy dispersion around $(0,\pi )$, but the
overall bandstructure is inconsistent with the experimental
one.  Several groups have also calculated the energy
dispersions of hole in the three bands model based on either
variational wavefunctions\cite{Nazarenko95,Vos} or
SCBA\cite{Starykh95} and found that the results reproduce
very well the experimental data.  These studies seem to
imply that the one band model is inferior to the multiband
model even for studying low energy properties of high-$T_c$
cuprates\cite{Vos}.  This is actually not true.  To
understand qualitatively the physics of a hole propagating
in an antiferromagnetic backgroun, the $t-J$ or $t-t^\prime
-J $ model might be sufficient.  However, to fit
quantitatively the result of a one band model with the
experimental data, the one band model used should include
all the terms which are deduced from the multiband Hubbard
model by eliminating highlying orbitals.  At least two of
those terms which are ignored in the previous study should
be considered, one is the $t^{\prime\prime}$-term (i.e.  the
next-next neighbor hopping term) and the other is the
three-site hopping term.  Both the $t^\prime$-term and
$t^{\prime\prime}$-term are originated from the effective
hopping (or wavefunction overlap) between two O $2p$
orbitals besides a Cu via the Cu $3d$ and $4s$
orbitals\cite{Wheatley}; the $t^\prime$-term is the hopping
between two nearest neighboring Oxygens, and the
$t^{\prime\prime}$-term is the hopping between two Oxygens
on the two sides of Cu.  $|t^{\prime\prime}|$ is generally
smaller than but certainly of the same order of magnitude as
$|t^\prime |$\cite{notes}.  The three site hopping term is
always present if the effective one band model is derived
from the one band Hubbard model\cite{Hirsch} or the
multiband Hubbard model\cite{Ogata}.  Both the
$t^{\prime\prime}$ and three-site hopping terms involve
hoppings on same sublattices and are affected very weakly by
antiferromagnetic correlations.  They may therefore have
substantial contributions to the quasiparticle dispersion.
Other long-range hopping terms involve the wavefunction
overlap of O orbitals not within the same unit cell and are
generally small and negligible.

In this paper we report our theoretical results for the energy 
dispersion in the one-band model with one hole in two
dimensions.  Both the Lanczos diagonalization and the SCBA
calculations are performed and compared.  We find that the
experimental data for ${\rm Sr_2CuO_2Cl_2}$ can be
quantitatively fitted by the $t-t^{\prime}-t^{\prime\prime}
- J$ model with the three-site hopping term within the
experimental error.

Let us first consider the $t-t^{\prime} -t^{\prime\prime} -
J$ model without the three-site hopping term.  The model
Hamiltonian is defined in the Hilbert subspace without
double occupied sites by
\begin{equation} 
 H= -\sum_{i \delta \sigma} (t_\delta
 c_{i\sigma}^\dagger c_{i+\delta\sigma} + H.c.)  + J
 \sum_{\langle ij\rangle } ({\bf S}_i\cdot {\bf S}_j -
 {1\over 4} n_in_j), \label{tjmodel} 
\end{equation} 
where $\langle \,\, \rangle$ refers nearest neighbors, and
$t_\delta =t$ for the nearest neighbor hopping ($\delta=
{\hat x}, \,{\hat y}$), $t^\prime$ for the next nearest
neighbor hopping ($\delta= {\hat x}\pm {\hat y}$), and
$t^{\prime \prime}$ for the next-next nearest neighbor
hopping ($\delta={2\hat x},\, {2\hat y}$).  The rest of
notation is standard.  For high-$T_c$ materials, both $t$
and $J$ vary in a certain energy scale, $t\approx 0.3
\sim 0.4 {\rm ev}$ and $J\approx 0.1 \sim 0.2 {\rm
ev}$, depending on compounds.  So far accurate
determinations for $t$ and $J$ from experiment are still not
available.  For concreteness in discussion, we take
$t=0.35$ev and $J=0.15$ev.  The quasiparticle bandwidth
evaluated from this set of values of $t$ and $J$ is roughly
the same as that for ${\rm Sr_2CuO_2Cl_2}$ when $t^\prime
=t^{\prime \prime}=0$.  In our calculation, we take
$t^\prime$ and $t^{\prime \prime}$ as free parameters, but
restrict $|t^{\prime \prime} | < |t^\prime | < J$.  If the
Hamiltonian (\ref{tjmodel}) is derived from the multiband
Hubbard model including the Cu $3d_{x^2-y^2}$ and $4s$ and O
$p_x$ and $p_y$ orbitals, then $t^\prime$ is 
negative (positive) for hole (electron) doped materials and
$t^{\prime \prime}$ has an opposite sign to
$t^\prime$\cite{Wheatley}.

For a one-hole system, we can do a Galilean transformation
to shift the origin of the frame of coordinates to the
position of hole\cite{Xiang91}.  If the hole momentum is
$k$, following the derivation of Ref.  \cite{Xiang91} it can
then be shown that the Hamiltonian (\ref{tjmodel}) is
equivalent to an effective Hamiltonian
\begin{eqnarray} 
H_h(k)&=& -\sum_{\delta\sigma} (t_\delta e^{-i(P -k)\cdot \delta}
c_{{\bf 0}\sigma}^\dagger c_{\delta\sigma} + H.c.)  \nonumber\\
&& + J
\sum_{\langle ij\rangle } ({\bf S}_i\cdot {\bf S}_j -
{1\over 4} n_in_j) \label{effective} 
\end{eqnarray}
where ${\bf 0}=(0,\,0)$ and $P= \sum_{k\sigma}
kc_{k \sigma}^\dagger c_{k\sigma}$ is the total momentum
operator.  In our exact diagonalization study, we solve this
effective Hamiltonian on finite size lattices with periodic
boundary condition (PBC), but allow $k$ to change
continuously (i.e.  $k$ can take all values allowed in an
infinite square lattice).

For each given $k$, we define the difference between the
lowest energy of the one-hole Hamiltonian (\ref{effective})
with total spin 1/2, $E_{N-1}(k)$, and the ground state
energy of the system without holes, $E_N$, as the coherent
hole quasiparticle energy $E(k)=E_N-E_{N-1}(k)$.  The
quasiparticle energy $E(k)$ such defined corresponds
directly to what measured in experiments, since the ground
state of ${\rm Sr_2CuO_2Cl_2}$ is a spin singlet and the
state with one electron removed from the ground state by
high energy photons has spin 1/2.  On finite size lattices,
some higher spin states may have lower energies than
$E_{N-1}(k)$ for certain values of $k$, so $E_{N-1}(k)$ may
not always be the minimum energy of $H_h(k)$.  However, in
thermodynamic limit we believe that $E_{N-1}(k)$ will either
be the minimum of $H_h(k)$ or differ infinitesmally small
from the minimum energy of $H_h(k)$ for all $k$. Thus 
alternatively $E(k)$ can also be defined as the difference
between the lowest eigeneigen of $H_h(k)$ and the ground
state energy without holes.  In thermodynamic limit the
above two definitions should give the same result for
$E(k)$.

Under the SCBA\cite{Schmitt-Rink88,Kane89}, the
quasiparticle energy is given by the position of coherence
peak at the bottom of the spectral function $A(k,\omega
)=-{\rm Im }G(k, \omega )/\pi$, and the single particle
Green's function $G(k, \omega )$ is determined by the
self-consistent equation:
\begin{equation}
G(k, \omega ) = {1\over \omega-\varepsilon_k - 
{1\over N}\sum_q \Gamma^2(k, q) 
G(k-q, \omega - \Omega_q)}, \label{Greens-function}
\end{equation}
where $\varepsilon_k =4t^\prime \cos k_x\cos k_y + 2
t^{\prime\prime} (\cos 2 k_x + \cos 2 k_y)$ and $\Gamma (k,
q)= 4t(\gamma_{k-q}u_q+ \gamma_k v_q)$ with $\Omega_q = 4 J
\sqrt{1-\gamma_q}$, $u_k=[(1-\gamma_k^2)^{-1/2}+1]^{1/2}/
\sqrt{2}$, $v_k = -{\rm sgn}(\gamma_k)
[(1-\gamma_k^2)^{-1/2} -1]^{1/2}/ \sqrt{2}$, and
$\gamma_k=(\cos k_x + \cos k_y)/2$.  From Eq.
(\ref{Greens-function}), it is straightforward to show that
$G(k, \omega )$ (and therefore the quasiparticle dispersion)
is symmetric under the reflection about the magnetic
Brilliou zone boundary (i.e.  $k_x+k_y=\pi$) since
$\varepsilon_k$ is symmetric under this transformation.
This symmetry is  purely due to the approximations
made in this approach; the original Hamiltonian
(\ref{tjmodel}) does not have this symmetry.

There are several approximations involved in the SCBA
calculation:  the linear spin-wave expansion and the neglect
of crossing diagrams.  These approximations ignore vertex
corrections and the hole distortion to the spin background
and do not guarantee the hard-core nature of the slave
fermions and the Schwinger bosons.  In the absence of the
$t^\prime$ and $t^{\prime\prime}$ terms, the contribution
from the two-loop crossing diagram to vertexs is exactly
zero\cite{Liu92}.  Thus the the vertex correction in the
SCBA calculation for the t-J model is small.  However, in
the presence of these terms, the two-loop crossing diagrams
are generally nonzero.

For the t-J model, the band structure of hole quasiparticles
has been extensively studied by many
groups\cite{Kane89,Liu92,Dagotta,Leung96}, and the agreement
between the finite size calculation\cite{Dagotta,Leung96}
and the SCBA calculation\cite{Liu92} is remarkably good.
The quasiparticle dispersion in this model shows many
interesting features which are completely different from
that in the ordinary metal.  Firstly, the band maximum
locates at $(\pm \pi /2, \pm \pi /2)$; secondly, the
effective mass is very anisotropic, it has a larger value
along the zone diagonal and a smaller value along the
direction perpendicular; and thirdly, the band width scales
with a certain power of the exchange energy $J$ ($\sim
J^{2/3}$ when $t\ll J$) rather than the hopping constant
$t$.

In the previous finite size calculations, the Hamiltonian
(\ref{tjmodel}) with PBC were generally used.  As the number
of $k$ points allowed in a finite lattice with PBC is
limited, calculations on large lattices are generally
required in order to obtain a complete picture of the
quasiparticle band structure.  However, if the Hamiltonian
(\ref{effective}) with PBC is diagonalized and the hole
momentum $k$ in (\ref{effective}) is taken as a free
parameter (this is eqivalent to diagonalizing the
Hamiltonian (\ref{tjmodel}) with twisted boundary conditions
as shown in Ref.  \cite{Xiang91} ), we find that even on a
relatively small lattice, such as $N=20$, a comprehensive
picture of the quasiparticle band structure can be obtained.

In Fig.  1a, our finite size diagonalization results for
$E(k)$ on $N=16$ and $20$ are shown and compared with the
SCBA ones for the $t-J$ model.  The agreement between the
finite lattice calculation and the SCBA calculation in the
whole Brilliou zone is good.  The finite size effect, as
revealed by the difference between the $N=16$ and $N=20$
results and the difference between the finite lattice result
and that of SCBA is small.  On finite lattices, the band
maxima locate not exactly at $(\pm\pi /2 , \pm\pi /2)$, but
they tend to move towards these points as $N$
increases\cite{Xiang91}.

\begin{figure}
\leavevmode
\epsfxsize=8cm
\epsfbox{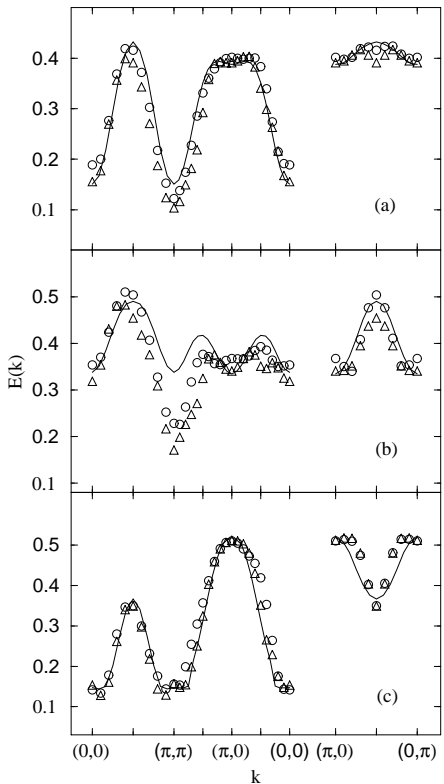}
\caption{Comparison of the quasiparticle dispersion relation $E(k)$ 
(unit: ev) 
obtained from the SCBA method on $24\times 24$ lattice 
(solid line) with that obtained 
from the finite size diagonalization method on $N=16$ (triangle) 
and $N=20$ (circle) lattices for the $t-t^\prime -J$ model 
with $t=0.35$ev, $J=0.15$ev and (a) $t^\prime =0$, (b) 
$t^\prime =-0.08$ev , and (c) $t^\prime =0.08$ev. 
\label{fig1}}
\end{figure}

The presence of the next nearest neighbor and the next-next
neighbor hopping terms changes largely the energy dispersion
of quasiholes.  Let us consider the contribution of the
$t^\prime$ term first.  When $t^\prime < 0$, as shown in 1b,
the energy dispersion along the line $(\pi ,0)-(0, \pi)$ is
enlarged.  Around $(\pi , 0)$, the dispersion is relatively 
small, but $E(k)$ falls much below the band maximum.  These
features are qualitatively consistent with the experimental
results for ${\rm Sr_2CuO_2Cl_2 }$\cite{Wells95}.  However,
the overall dispersion of $E(k)$ (except at the
vicinity of $(\pi ,\pi )$) is largely reduced by the
$t^\prime$ term.  If fitting the experimental data of ${\rm
Sr_2CuO_2Cl_2}$ using the $t-t^\prime -J$ model, we find
that a rather large $J$ value, about twice as large
as what has been found in experiments, is needed.

When $t^\prime > 0$, as shown in Fig.  1c, $E(k)$ behaves
quite differently from the case $t^\prime <0$:  (1) In this
case the band maxima locate at $(\pi ,0)$ and its equivalent
points; (2) the bandwidth is enlarged compared with the
$t-J$ model ; (3) $E(k)$ becomes more symmetric about the
magnetic Brillion zone boundary and shows a large dispersion
along the line $(\pi ,\pi )- (\pi ,0) - (0,0)$; (4) The
agreement between the SCBA and the finite size calculations
is fairly good and the finite size effect is even smaller
than that for the $t-J$ model.  However, as $t^{\prime}$
gets larger, there is no well defined coherent peak at the
bottom of quasiparicle spectrum when $k$ is in the vicinity
of $(0,0)$ in the SCBA.  So far no ARPES data are available
for electron doped cuprates for comparison.  If we believe
that the asymmetry between the electron and hole doped
high-$T_c$ materials is mainly due to the next nearest
neighbor hopping term, then the above features of $E(k)$
should in principle be observable in the ARPES spectrum in
electron doped materials.

For the $t-t^\prime -J$ model with a negative $t^\prime$
term, the overall agreement between the finite size
calculation and the SCBA calculation is not as good as for
the $t^\prime \ge 0$ cases.  This disagreement is probably
due to the linear spin-wave approximation in the SCBA
approach which ignores the influence of holes to the quantum
antiferromagnetic Neel state.  A positive (negative)
$t^\prime$ term can enhance (reduce) the antiferromagnetic
correlations in the one band model\cite{Asymmetry}.  So in
the presence of a negative $t^\prime$ term, the distortion
of a hole to the quantum antiferromagnetic Neel state is
enlarged.  In this case to obtain a more quantitative
description for the quasiparticle dispersion, exact
diagonalizations on larger lattices are needed.

\vspace{-3cm}
\begin{figure}
\leavevmode
\epsfxsize=6cm
\epsfbox{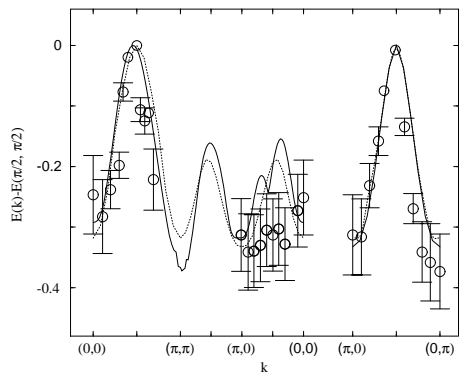}
\caption{Comparison of the quasiparticle energy 
$E(k)-E(\pi /2, \pi /2)$ (unit: ev) in 
${\rm Sr_2CuO_2Cl_2}$ (circle) with the corresponding 
results for the $t-t^\prime -t^{\prime\prime}-J$ model 
obtained from the exact diagonalization 
with $N=20$ (solid line) and the SCBA approach 
(dashed line). $t=0.35$ev, $t^\prime =-0.12$ev, 
$t^{\prime\prime} =0.08$ev, and $J=0.15$ev. 
\label{fig2}}
\end{figure}

The above discussion indicates that the bandwidth of the
$t-t^\prime -J $ model is too small compared with the
experimental one for ${\rm Sr_2CuO_2Cl_2 }$ within the
physically reasonable region of parameters, consistent 
with the previous study\cite{Nazarenko95}.   
A finite $t^{\prime\prime}$ term, however, can change this
situration significantly.  The contribution of the
$t^{\prime\prime}$ term to $E(k)$ is similar to a
$-t^\prime$ term, for  example for the
$t-t^{\prime\prime} - J$ model ($t^\prime =0$) the
band maximum locates at $(\pi /2 , \pi /2)$ if
$t^{\prime\prime} > 0$ or $(\pi ,0 )$ if $t^{\prime\prime} <
0$, but the overall energy dispersion is
always enlarged by the $t^{\prime\prime}$ term.  

In Fig.  (\ref{fig2}) the quasiparticle dispersions for the
$t-t^\prime -t^{\prime\prime}-J$ model are shown and
compared with the experimental data for ${\rm Sr_2CuO_2Cl_2
}$.  Along two diagonal lines, $(0,0)-(\pi ,\pi )$ and $(\pi,
0)-(0, \pi )$, the finite size effect is small and the
agreement between the experimental data and our calculation
is very good.  Compared with Fig.  1b, we find that the
bandwidth is largely enhanced by even a small
$t^{\prime\prime}$-term.  Along the line $(\pi \pi )-(\pi ,
0)-(0, 0)$, the energy dispersion obtained from both the
exact diagonalization and SCBA calculations is larger than
the experimental one, and the agreement between the exact
diagonalization and the SCBA results along this line is also
not as good as along two diagonal lines.  A sub-peak appears
in the curve of the exact diagonalization result on the line 
$(\pi ,0)-(0,0)$.  This is purely due to the finite size
effect.

\vspace{-3cm}
\begin{figure}
\leavevmode
\epsfxsize=6cm
\epsfbox{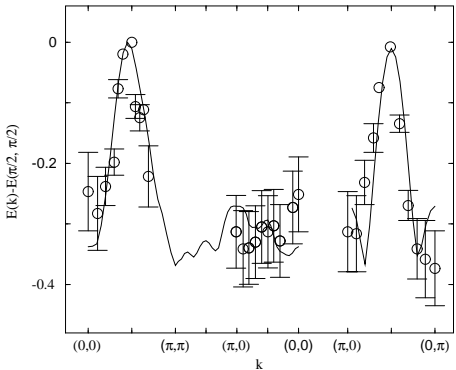}
\caption{Comparison between the quasiparticle energy 
$E(k)-E(\pi /2, \pi /2)$ (unit: ev) of the $t-t^\prime 
-t^{\prime\prime}-J$ model with the three-site 
hopping term on a $N=20$ lattice (curve) and that of 
${\rm Sr_2CuO_2Cl_2}$. $t=0.35$ev, $t^\prime =-0.12$ev, 
$t^{\prime\prime} =0.08$ev, and $J=0.15$ev. 
\label{fig3}}
\end{figure}

Now let us consider the three-site hopping term.  The
Hamiltonian for the three-site hopping term is given
by\cite{Hirsch,Ogata} 
\begin{equation} 
H_{3-site}= {J\over
4} \sum_{\langle ij\rangle \not= \langle ik \rangle \sigma}
(c_{i{\bar\sigma}}^\dagger c_{i\sigma} c_{j\sigma}^\dagger
c_{k {\bar\sigma}}-n_{i{\bar\sigma}}c_{j\sigma}^\dagger c_{k
\sigma}).  
\end{equation} 
This term describes an effective
hopping of a hole to one of its next or next-next neighbor
sites by exchanging spins with another hole on its nearest
neighbor site.  From the calculation we find that this term
has very weak effect on the quasiparticle dispersion along
two diagonal lines.  However, it has a relatively larger
effect on $E(k)$ when $k$ varies along $(\pi ,\pi )- (\pi , 0) -
(0, 0)$.  It suppresses the dispersion of $E(k)$ on $(\pi
,\pi )- (\pi , 0) - (0, 0)$ and therefore improves the
agreement between the theoretical result and the
experimental data.  Fig.  (\ref{fig3}) compares the energy
dispersion of the $t-t^\prime-t^{\prime\prime} -J$ model with 
the three-site term  on
a $N=20$ lattice with that of ${\rm Sr_2CuO_2Cl_2 }$.  It is
easy to see that the one-band $t-t^\prime -t^{\prime \prime}
-J $ model with the three-site hopping term gives a good
account for the quasiparticle dispersion in ${\rm
Sr_2CuO_2Cl_2 }$.  This result is consistent with the
previous studies for the quasiparticle dispersion based upon
the multiband Hubbard model\cite{Nazarenko95,Vos,Starykh95}.
It implies that the the multiband Hubbard model is indeed 
equivalent to a one-band model in describing low energy
excitations of high-$T_c$ cuprates.

In conclusion, we have studied the energy dispersion of hole
quasiparticles in an antiferromagnetically correlated
background using the exact diagonalization and the SCBA
methods.  In the exact diagonalization study, we first
derive an effective Hamiltonian for the one-hole t-J model
using the Galilean transformation for each given hole
momentum $k$ (which can take all allowed values in an
infinite lattice), and then diagonalize this effective
Hamiltonian on finite lattices with PBC.  As no limitation
for $k$ values, this allows us to have a comprehensive study
for the band structure of quasiparticles with the finite
size diagonalization even on a $N=20$ lattice.  We find that
the finite size diagonalization results agree well with the
SCBA ones, especially when $t^\prime \ge 0$.  For hole doped
high-$T_c$ compounds, the band maxima locate at $(\pm \pi /2
, \pm \pi /2)$ in thermodynamic limit.  However, for
electron doped compounds, the band maxima locate at $(\pi,
0)$ and its equivalent points.  The
experimental data for ${\rm Sr_2 Cu O_2 Cl_2}$ can be
quantitatively understood from the one band $t-t^\prime
-t^{\prime \prime}- J $ model with the three-site hopping
term.


\begin{references}
\bibitem{notes} More precious calculation for the values
of $t^\prime$ and $t^{\prime\prime}$ needs consider the
orthogonality of the Wannier representation of O $2p$
bonding and nonbonding orbitals.  Direct O-O wavefunction
overlap may have also a small contribution to $t^\prime$.
\end{references}
\end{document}